\documentclass[11pt]{article}
\usepackage{latexsym}  
\usepackage{amssymb}
\usepackage{graphicx}
\usepackage{amsmath}

\newcommand{\be}{\begin{equation}}
\newcommand{\ee}{\end{equation}}
\newcommand{\bea}{\begin{eqnarray}}
\newcommand{\eea}{\end{eqnarray}}
\newcommand{\bref}[1]{(\ref{#1})}

\begin{document}
\begin{titlepage}
\begin{flushright}
May 10, 2014
\end{flushright}
\vspace{4\baselineskip}
\begin{center}
{\Large\bf  Simple neutrino mass matrix with only two free parameters}
\end{center}
\vspace{1cm}
\begin{center}
{\large Hiroyuki Nishiura$^{a,}$
\footnote{E-mail:nishiura@is.oit.ac.jp}}
and
{\large Takeshi Fukuyama$^{b,}$
\footnote{E-mail:fukuyama@se.ritsumei.ac.jp}}
\end{center}
\vspace{0.2cm}
\begin{center}
${}^{a}$ {\small \it Faculty of Information Science and Technology, 
Osaka Institute of Technology,\\ Hirakata, Osaka 573-0196, Japan}\\[.2cm]

${}^{b} $ {\small \it Research Center for Nuclear Physics (RCNP),
Osaka University, Ibaraki, Osaka, 567-0047, Japan}

\vskip 10mm
\end{center}
\vskip 10mm
\begin{abstract}
A simple form of neutrino mass matrix which has only two free parameters 
is proposed from a phenomenological point of view.
Using this mass matrix, we succeed to reproduce all the observed values for the MNS lepton mixing angles 
and the neutrino mass squared difference ratio.
Our model also predicts $\delta_{\nu}= 155^\circ$ for the Dirac CP 
violating phase in the lepton sector and the effective neutrino mass $\langle m \rangle=6.3\times 10^{-3}$ eV in the neutrinoless double beta decay.
\end{abstract}
PCAC numbers:  
  14.60.Pq,  
  12.60.-i, 
\end{titlepage}
The mixing angles of Maki-Nakagawa-Sakata (MNS) lepton mixing matrix \cite{MNS} \cite{Pontecorvo}
and the neutrino mass squared differences have been experimentally determined 
from the neutrino oscillation experiments.
These lepton mixing angles indicate that neutrino flavors mix nearly maximally unlike quark mixings.
On the other hand, the absolute values of the neutrino masses $m_i$, the Dirac, and the Majorana CP violating phases in the MNS lepton mixing matrix are still remained undetermined.
The origin of nearly maximally mixing are investigated from the point of neutrino mass matrix structure using a $\mu-\tau$ symmetry \cite{F-N} which predicts $\theta_{23}=\pi/4$ and $\theta_{13}=0$. The recent finding \cite{theta13} \cite{Daya-Bay} \cite{RENO} \cite{Daya-Bay2} of a relatively large $\theta_{13}$ forces us to consider its origin and model extensions.   

The rather large $\theta_{13}$ opens the possibility of CP violation also in lepton sector. 
In this stage it is suitable to consider a simple model incorporating CP violationg phase. 
There we should not be too much constrained by the symmetries of preceding real matrix models like the $\mu-\tau$ or the other symmetries. 
Indeed, neutrino mass matrix in many realistic GUTs does not match well with such symmetry \cite{Fukuyama-Okada}. 
In constructing model in this paper, we do not consider hierarchical structure since hierarchy needs some symmetry to protect it.

In this paper, we propose a simple 2 parameter complex mass matrix for Majorana neutrinos. 
The aim of the present paper is to propose a model minimizing the number of free parameters as far as possible. 
Its form is
\be
M_\nu=k_\nu \left[\left(
\begin{array}{ccc}
-1 & 1 & 1\\
1 & 1 & 1\\
1 & 1 & -1
\end{array}
\right)
+
\left(
\begin{array}{ccc}
0 & 0 & 0\\
0 & A^* i & A\\
0 & A & A
\end{array}
\right)
\right]. \label{mass_matrix_model}
\ee
Here, $A$ is a complex parameter parameterized by two real parameters $x$ and $y$ as follows, 
\bea
A&=&x + yi,\\
A^* i&=&y + xi.
\eea
Note that $A^* i$ is obtained from $A$ by exchanging the real part with the imaginary part of it.
The neutrino mass matrix \bref{mass_matrix_model} consists of two terms. The first term are pseudo democratic mass matrix, while the  second term is also semi democratic only in the  second and third generations. 
Since we are interested only in the mass ratios and mixings, we neglect the overall constant $k_\nu$ in the following discussions.

The Majorana neutrino mass matrix $M_\nu$ is diagonalized as
\be
M_\nu=U\left(
\begin{array}{ccc}
m_{\nu 1}&0&0\\
0&m_{\nu 2}&0\\
0&0&m_{\nu 3}
\end{array}
\right)U^T,
\ee
where $m_{\nu i}$ are neutrino masses.
We assume that the mass matrix for the charged leptons is diagonal, so that the $U$ is the MNS lepton mixing matrix itself, which is represented in the standard form,
\bea
U_{std}&=& \mbox{diag} (e^{i \alpha_1^e}, e^{i \alpha_2^e}, e^{i \alpha_3}) U\\ \nonumber
&=&\left(
\begin{array}{ccc}
c_{12}c_{13}&s_{12}c_{13}&s_{13}e^{-i\delta_\nu }\\
-s_{12}c_{23}-c_{12}s_{13}s_{23}e^{i\delta_\nu }&
c_{12}c_{23}-s_{12}s_{13}s_{23}e^{i\delta_\nu}&c_{13}s_{23}\\
s_{12}s_{23}-c_{12}s_{13}c_{23}e^{i\delta_\nu }&
-c_{12}s_{23}-s_{12}s_{13}c_{23}e^{i\delta_\nu }&c_{13}c_{23}\\
\end{array}
\right)
\left(
\begin{array}{ccc}
1 & 0 & 0\\
0 & e^{i\beta} & 0\\
0 & 0 & e^{i\gamma}\\
\end{array}
\right). \label{mixing}
\eea
Here, $e^{i \alpha_i^e}$ comes from the rephasing in the charged lepton fields to make the choice of phase convention, and $c_{ij}=\cos\theta_{ij}$ and $s_{ij}=\sin\theta_{ij}$ with $\theta_{ij}$ being lepton mixing angles. 
The $\delta_\nu $ is the Dirac CP violating phase and $\beta $ and $\gamma $ are the Majorana CP violating phases.

In the mass matrix model given in (\ref{mass_matrix_model}), the MNS mixing parameters ($\sin^2 2 \theta_{12}$, 
$\sin^2 2 \theta_{23}$, and $\sin^2 2 \theta_{13}$) and the neutrino mass squared difference ratio ($R_\nu \equiv {\Delta m_{21}^2}/{\Delta m_{32}^2}$) are functions of only two free parameters $x$ and $y$.  
On the other hand, the observed values of the MNS mixing parameters and $R_{\nu}$ are \cite{PDG}
\be
\begin{array}{l}
\sin^2 2\theta_{12} = 0.857 \pm 0.024, \\
\sin^2 2\theta_{13} = 0.095 \pm 0.010, \\
\sin^2 2\theta_{23} >0.95,  \label{exptdata1}
\end{array}
\ee
\be
R_{\nu} \equiv \frac{\Delta m_{21}^2}{\Delta m_{32}^2}
=\frac{m_{\nu2}^2 -m_{\nu1}^2}{m_{\nu3}^2 -m_{\nu2}^2}
=\frac{(7.50\pm 0.20) \times 10^{-5}\ {\rm eV}^2}{
(2.32^{+0.12}_{-0.08}) \times 10^{-3}\ {\rm eV}^2} = 
(3.23^{+0.14}_{-0.19} ) \times 10^{-2} .\label{exptdata2}
\ee
It is shown later that our mass matrix model \bref{mass_matrix_model} well reproduces the observed values 
for the case of normal neutrino mass hierarchy.

Using Eqs.\bref{exptdata1} and \bref{exptdata2}, we draw in Fig.~1 the contour curves in the $(x, y)$ parameter plane corresponding to the observed values of the MNS mixing parameters ($\sin^2 2 \theta_{12}$, 
$\sin^2 2 \theta_{23}$, and $\sin^2 2 \theta_{13}$) and the neutrino mass squared difference ratio $R_\nu$. 
We find that the parameter set around $(x, y)=(0.725, 3.43)$  
indicated by a star ($\star$) in Fig.~1 is 
consistent with all the observed values.
Namely all the observed three MNS mixing angles and the neutrino 
mass squared difference ratio are well reproduced if we take values of $x$ and $y$ as 
\be
 (x, y)=(0.725, 3.43). \label{abvalue}
\ee
In order to see the sensitivity to $x$ and $y$, we show $x$ and $y$ dependences of the MNS mixing parameters 
$\sin^2 2\theta_{12}$, $\sin^2 2\theta_{23}$, $\sin^2 2\theta_{12}$, 
and $10R_{\nu}$ in Fig.~2 and Fig.~3, respectively. 
It is found that $10R_{\nu}$ and $\sin^2 2 \theta_{12}$ are very sensitive to $x$ and $y$. 

Let us present the predictions from our model.
By taking the values for the free parameters $x=0.725$ and $y=3.43$, we predicts
\bea
\sin^2 2\theta_{12} = 0.881, \\
\sin^2 2\theta_{13} = 0.0925, \\
\sin^2 2\theta_{23} = 0.983, \\
R_\nu = 0.0307,\\
\delta_\nu = 155^\circ (= 0.863 \pi),\\
\beta = 172^\circ,\\
\gamma = 22.7^\circ\, .
\eea
This is also consistent with the recent global best fit values \cite{Fogli} except for $\delta_\nu$, their best fit $\delta_\nu=1.39\pi$ and $\delta_\nu~\mbox{in}~1\sigma=(1.12-1.72)\pi$.

We can also predict the absolute neutrino masses, for the parameters given by \bref{abvalue}, 
as follows
\be
m_{\nu 1} \simeq 0.0031\ {\rm eV}, \ \ m_{\nu 2} \simeq 0.0092 \ {\rm eV}, 
\ \ m_{\nu 3} \simeq 0.050 \ {\rm eV}  ,
\ee
by using the input value \cite{MINOS13}
$\Delta m^2_{32}\simeq 0.00241$ eV$^2$.
Our model also predicts the effective Majorana neutrino mass \cite{Doi1981} 
$\langle m \rangle$ 
in the neutrinoless double beta decay as
\be
\langle m \rangle =\left|m_{\nu 1} (U_{e1})^2 +m_{\nu 2} 
(U_{e2})^2 +m_{\nu 3} (U_{e3})^2\right| 
\simeq 6.3 \times 10^{-3}\ {\rm eV}.
\ee

In conclusion, we have proposed the phenomenological semi-democratic neutrino mass matrix 
with only two free parameters. 
Using this mass matrix we have been able to reproduce all the neutrino oscillation data consistently and
predict three unknown CP violating phases.

\section*{Acknowledgements}
The work of T.F.\ is supported in part by the Grant-in-Aid for Science Research
from the Ministry of Education, Science and Culture of Japan
(No.~26247036).

\newpage
\begin{figure}[ht]
\begin{center}
\includegraphics[height=.4\textheight]{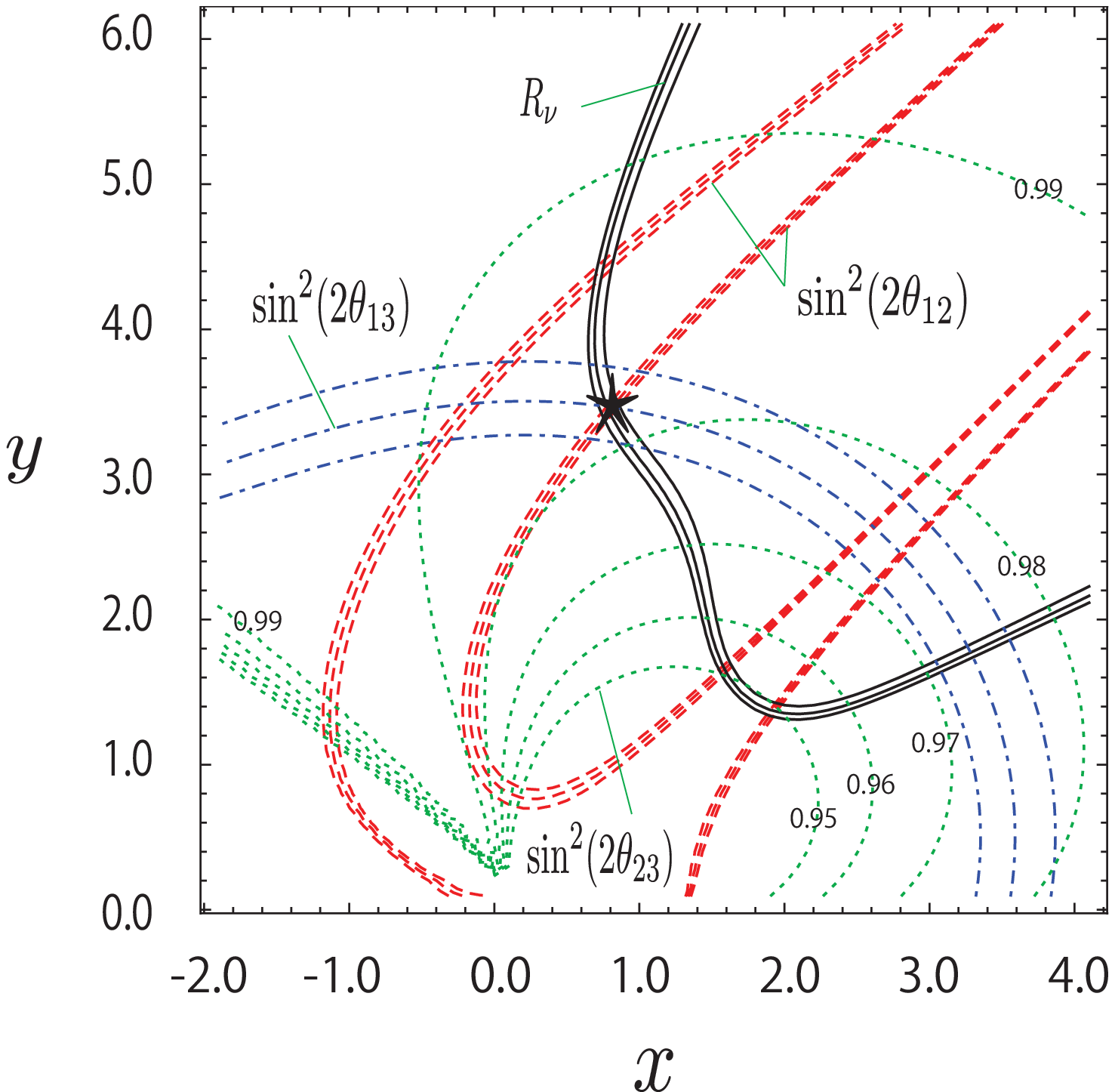}
\end{center}  
  \caption{Contour plots in the $(x, y)$ parameter plane for the lepton mixing parameters 
$\sin^2 2\theta_{12}$(dashed), 
$\sin^2 2\theta_{23}$(dotted), $\sin^2 2\theta_{13}$(dotdashed), and the neutrino 
mass squared difference ratio $R_\nu$(solid).  Contour curves for the observed (center, upper, and lower) values given in \bref{exptdata1} and \bref{exptdata2} are drawn. As for the $\sin^2 2\theta_{23}$, contour curves for $\sin^2 2\theta_{23}=0.95, 0.96, 0.97, 0.98, 0.99$ are drawn.
}\label{fig1}
\end{figure}
\begin{figure}[ht]
\begin{center}
\includegraphics[height=.3\textheight]{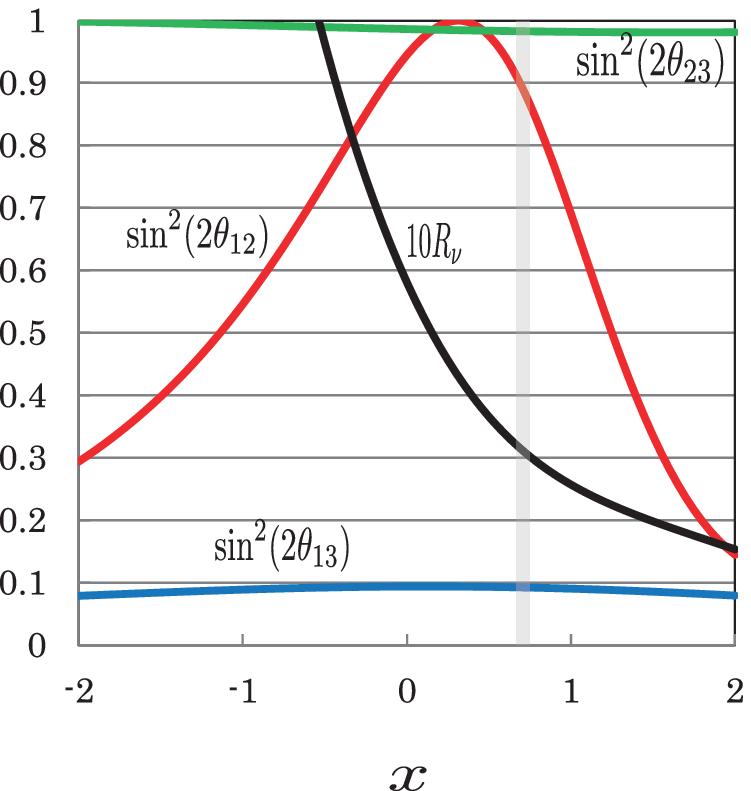}
\end{center}  
  \caption{$x$ dependence of $\sin^2 2\theta_{12}$, 
$\sin^2 2\theta_{23}$, $\sin^2 2\theta_{13}$, and $10R_\nu$. 
We have drawn curves of them as functions of $x$ with taking $y=3.43$.  
}\label{fig2}
\end{figure}
\begin{figure}[ht]
\begin{center}
\includegraphics[height=.3\textheight]{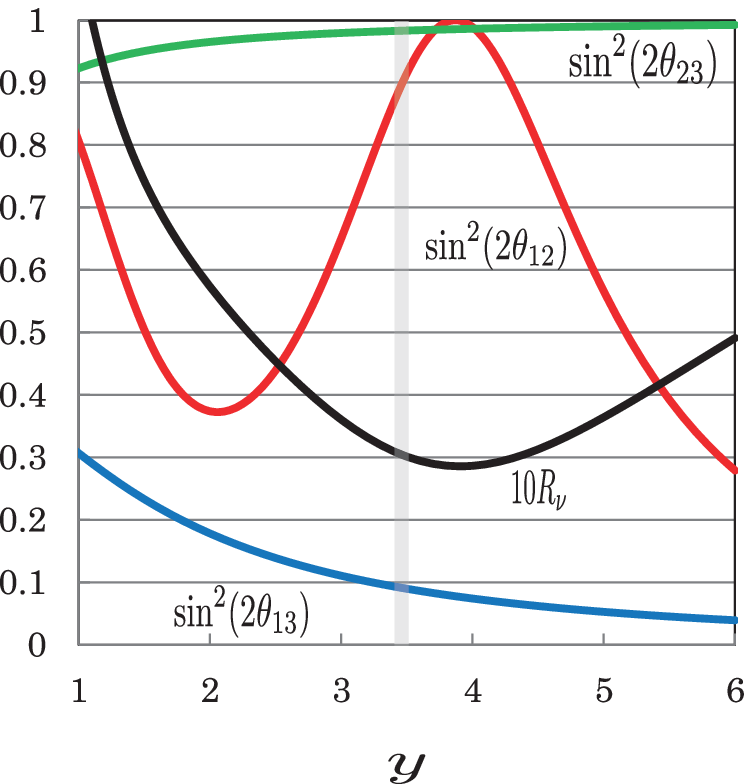}
\end{center}  
  \caption{$y$ dependence of $\sin^2 2\theta_{12}$, 
$\sin^2 2\theta_{23}$, $\sin^2 2\theta_{13}$, and $10R_\nu$. 
We have drawn curves of them as functions of $y$ with taking $x=0.725$.  
}\label{fig3}
\end{figure}

\end{document}